%% file: root.tex
\documentclass[letterpaper, 10 pt, conference]{ieeeconf}  % Comment this line out if you need a4paper

\IEEEoverridecommandlockouts                              % This command is only needed if 
\usepackage{subfiles}
\usepackage{amsmath}
\usepackage{graphicx}
\usepackage{multirow}
\usepackage{subcaption}
\usepackage{array}

\usepackage[natbib, bibstyle=ieee]{biblatex}
\title{\LARGE \bf
AI-assisted Automatic Jump Detection and Height Estimation in Volleyball Using a Waist-worn IMU
}

\author{Weiyi Xu$^{1}$, Chunzhuo Wang$^{1}$ Meng Shang$^{1}$, Camilla De Bleecker$^{2}$, \\ Maria Torres Vega$^{1}$, Jos Vanrenterghem$^{3}$, Bart Vanrumste$^{1}$ % <-this % stops a space
\thanks{*This work was in part by China Scholarship Council (CSC) under grant No. 202407960007, and in part by Flanders AI Research program.}% <-this % stops a space
\thanks{$^{1}$ W. Xu, C. Wang, M. Shang, M. Torres Vega and B. Vanrumste are with e-Media Research Lab, KU Leuven, Leuven, Belgium 
    {\tt\small weiyi.xu@kuleuven.be. 
    }}%
\thanks{$^{2}$
        C. De Bleecker is with Department of Rehabilitation Sciences, Ghent University, Ghent, Belgium.
        }%
\thanks{$^{3}$
        J. Vanrenterghem is with Department of Rehabilitation Sciences, KU Leuven, Leuven, Belgium.
        }%
}

\begin{document}

\maketitle
\thispagestyle{empty}
\pagestyle{empty}

%%%%%%%%%%%%%%%%%%%%%%%%%%%%%%%%%%%%%%%%%%%%%%%%%%%%%%%%%%%%%%%%%%%%
\begin{abstract}
The physical load of jumps plays a critical role in injury prevention for volleyball players. However, manual video analysis of jump activities is time-intensive and costly, requiring significant effort and expensive hardware setups. The advent of the inertial measurement unit (IMU) and machine learning algorithms offers a convenient and efficient alternative. Despite this, previous research has largely focused on either jump classification or physical load estimation, leaving a gap in integrated solutions. This study aims to present a pipeline to automatically detect jumps and predict heights using data from a waist-worn IMU. The pipeline leverages a Multi-Stage Temporal Convolutional Network (MS-TCN) to detect jump segments in time-series data and classify the specific jump category. Subsequently, jump heights are estimated using three downstream regression machine learning models based on the identified segments. Our method is verified on a dataset comprising 10 players and 337 jumps. Compared to the result of VERT in height estimation (R-squared = $-1.53$), a commercial device commonly used in jump landing tasks, our method not only accurately identifies jump activities and their specific types (F1-score = $0.90$) but also demonstrates superior performance in height prediction (R-squared = $0.50$). This integrated solution offers a promising tool for monitoring physical load and mitigating injury risk in volleyball players.
\newline

\end{abstract}

%%%%%%%%%%%%%%%%%%%%%%%%%%%%%%%%%%%%%%%%%%%%%%%%%%%%%%%%%%%%%%%%%%%%%%%%%%%%%%%%
\section{INTRODUCTION}
\subfile{section/Introduction.tex}

\section{Methods}
\subfile{section/Methods.tex}

\section{Experiments and Discussion}
\subfile{section/Experiments_and_discussion.tex}

\section{CONCLUSION}

We propose a DL based pipeline to identify jumps and predict heights using a waist-worn IMU during volleyball training and games. The integration approach can automatically detect jumps during the entire session with accurate category labeling and height estimation, outperforming the commercial VERT device. Future work can consider more direct musculoskeletal load intensity measurements, such as patellar tendon force, to better quantify the relationship between the volleyball jumps and impact loads.


%%%%%%%%%%%%%%%%%%%%%%%%%%%%%%%%%%%%%%%%%%%%%%%%%%%%%%%%%%%%%%%%%%%%%%%%%%%%%%%%

%%%%%%%%%%%%%%%%%%%%%%%%%%%%%%%%%%%%%%%%%%%%%%%%%%%%%%%%%%%%%%%%%%%%%%%%%%%%%%%%

%%%%%%%%%%%%%%%%%%%%%%%%%%%%%%%%%%%%%%%%%%%%%%%%%%%%%%%%%%%%%%%%%%%%%%%%%%%%%%%%

% \section*{ACKNOWLEDGMENT}

% The preferred spelling of the word ÒacknowledgmentÓ in America is without an ÒeÓ after the ÒgÓ. Avoid the stilted expression, ÒOne of us (R. B. G.) thanks . . .Ó  Instead, try ÒR. B. G. thanksÓ. Put sponsor acknowledgments in the unnumbered footnote on the first page.

%%%%%%%%%%%%%%%%%%%%%%%%%%%%%%%%%%%%%%%%%%%%%%%%%%%%%%%%%%%%%%%%%%%%%%%%%%%%%%%%
\printbibliography

% \bibliographystyle{IEEEtran}
% \bibliography{template/ref}

\end{document}

%% file: section/Introduction.tex
Jumping is one of the most important elements in volleyball, serving as a key indicator of athletic performance ~\cite{pawlik2023influence}. Meanwhile, intensive jump tasks pose significant risk to athletes' physical health. According to \cite{migliorini2019injuries}, frequent jumps can lead to injuries such as patellar tendonitis, caused by overuse of the knee joint during jump activities. In addition to common vertical jumps seen in other sports, volleyball demands sport-specific jumps, such as block jumps and attack jumps \cite{sattler2012vertical}. Performing such variety in jumps may lead to injuries caused by excessive stress on key musculoskeletal structures \cite{migliorini2019injuries}. Therefore, accurately assessing the physical impact of volleyball jumps requires analyzing both the frequency and type of jumps. Apart from these factors, to further evaluate the potential physical load, existing research has used jump height as an indicator of the magnitude of forces on body tissues \cite{jones2016strength, kons2018vertical}.

Video-based analysis is the most widely used method for evaluating performance in volleyball and other sports \cite{takahashi2016robust, johnston2018video}. However, achieving accurate observations often requires multiple professional cameras, which can be costly. It is also a labor-intensive process due to the significant workload involved in video analysis. Wearable inertial measurement units (IMU) provides a more efficient approach to monitoring load in sports \cite{rana2020wearable}. IMUs are applied in various sports-related tasks, such as activity classification in basketball \cite{ma2018basketball}, or jumping height estimation in beach volleyball (calculated by flight time) \cite{schleitzer2022development}. In addition to the advanced research, commercial device such as VERT, is widely used for jump counting and load monitoring \cite{brooks2024quantifying}, although the proprietary algorithms is not publicly available.                                                                         

Unlike traditional method that requires signal processing by prior knowledge in sports-related kinetics and kinematics \cite{schleitzer2022development}, Machine Learning (ML) and Deep Learning (DL) models provide an efficient approach for impact activity analysis from IMU signals. \citet{mascia2023machine} selected a Multilayer Perceptron model (MLP) and collected signals via a smartphone to predict jump heights. The result of root mean squared distance is at 4 cm, although this work is limited on estimating heights only for Counter Movement Jump (CMJ). Meanwhile, this work collected signals in separate jumping episodes and skipped the phase of jump detection. On the other hand, \citet{kautz2017activity} proposed a DL based method for jump detection in beach volleyball and achieves an overall accuracy of $83.2\%$ without investigation into the physical load assessment. However, both the jumping counts and corresponding heights are essential to monitor the comprehensive load profile. To the best of our understanding, no previous research has investigated ML based integrated pipeline to efficiently automate the load monitoring process for volleyball players.

In summary, this study proposes a novel pipeline leveraging ML models for jump height estimation using IMU data in volleyball training and games. It comprises two sequential components: jump detection and height estimation. The jump height is estimated based on the detected jump relevant segments from the classification model. This integrated approach enables to automatically calculate potential physical load by observing intensive jump tasks for a volleyball player by a waist-worn IMU during the entire recording session.

%% file: section/Methods.tex
\begin{figure*}[htpb]
    \centering
    \includegraphics[width=\linewidth]{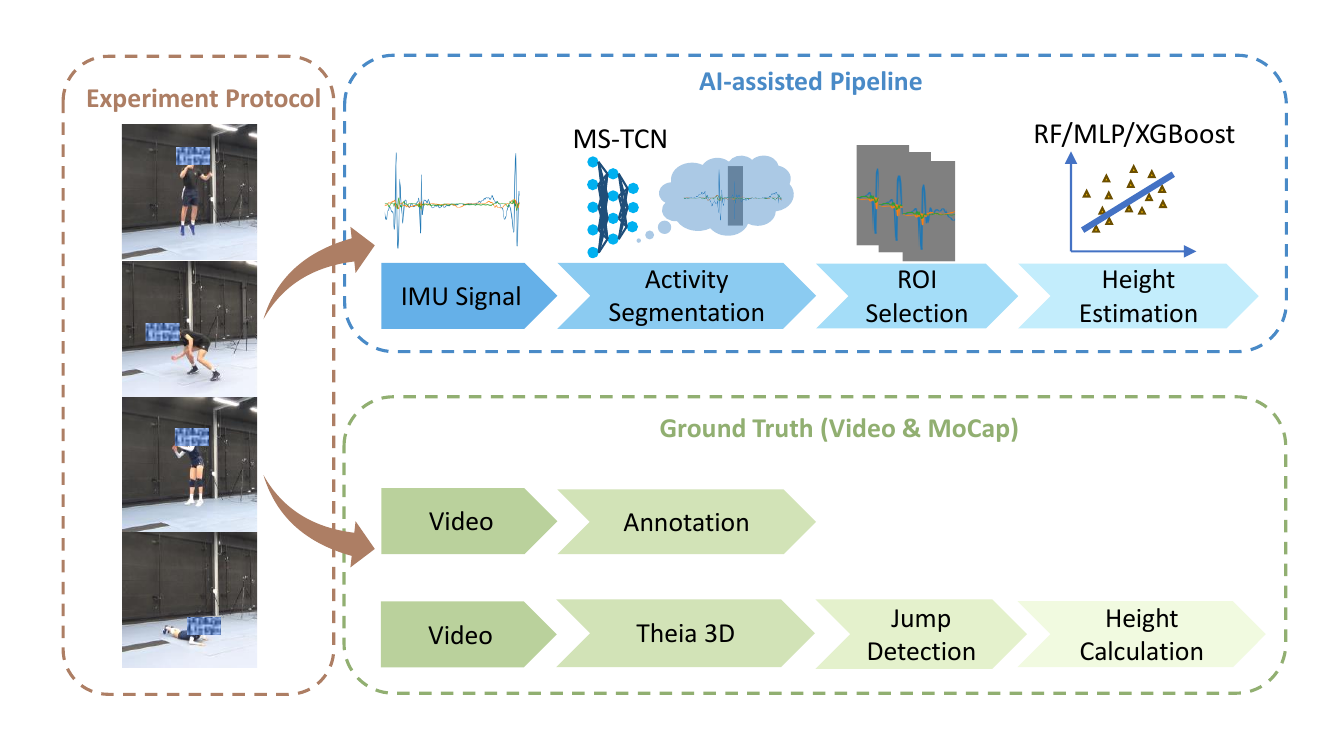}
    \caption{Overview of the framework of this study. The data is collected using a waist-mounted IMU following by the experiment protocol that involves multiple jumps demanded in volleyball games. Our proposed AI-assisted pipeline takes the raw input of signals for jump detection and classification using a DL model. The identified segments from the classification report are used to prepare handcrafted features as the input to several ML models for height estimation. The ground truth is obtained by video annotation and motion capture (MoCap) software Theia 3D \protect\footnotemark[1].}
    \label{fig:2-1}
\end{figure*}
\footnotetext[1]{https://www.theiamarkerless.com/}

The complete framework for jump height estimation is illustrated in Fig. \ref{fig:2-1}. The process begins with raw IMU signals. As in our proposed pipeline, a DL model is employed to predict a label for each sample within the time-series data in the first stage. Using the predicted continuous segments classified as certain jump types, the regression models on the second stage can predict the corresponding heights.

\subsection{Activity segmentation}
To ensure high resolution in the classification report, we adopt a sample-wise classification method. Specifically, for each participant, $I = \{I_1, I_2, \ldots, I_N\}$ represents the input IMU signals, where $I_n \in R^{C}$ ($1 \le n \le N$), $N$ denotes the length of the time series and $C$ denotes the number of IMU channels. The objective of the sample-wise classification model is to provide an output as $O = \{O_1, O_2, \ldots, O_N\}$, where $O_n \in R^J$ represents the probability of $n-$th timestamp on each jump categories, and $J$ equals to the number of jump types (including non-jump type). The model processes the entire time series of IMU signals, and aims to give a prediction on each timestamp.

We selected the multi-stage temporal convolutional network (MS-TCN) as the classification model. \citet{lea2017temporal} proposed the temporal convolutional network (TCN) for segmentation classification task. By introducing dilated factor to increase the receptive field, TCN can capture the temporal features within the long range of time series without significantly increasing the model parameters. We refer to this fundamental TCN network as the single-stage TCN (SS-TCN). Multi-stage TCN (MS-TCN) is a stack of (SS-TCN) \cite{farha2019ms}, that attempts to train each subsequent SS-TCN by using the predictions from the previous stage and refine the results.

\subsection{ROI selection}
Based on the classification report, consecutive samples categorized as specific jump types are considered as jump-related segments. Considering the discrepant length of the identified segments, it is crucial to ensure the consistent input dimensions while preserving essential information for the height estimation task. A large size of the Region of Interest (ROI) is necessary to incorporate not only the jump landing phase but also preceding and following signals. Previous research \cite{mascia2023machine} has shown that multiple peak signals can occur before or after the actual flight phase of a jump. Additionally, using a larger ROI helps mitigate potential misalignment caused by annotation errors and discrepancies in sampling frequencies across devices.

\begin{figure}[htb]
    \centering
    \includegraphics[width=\linewidth]{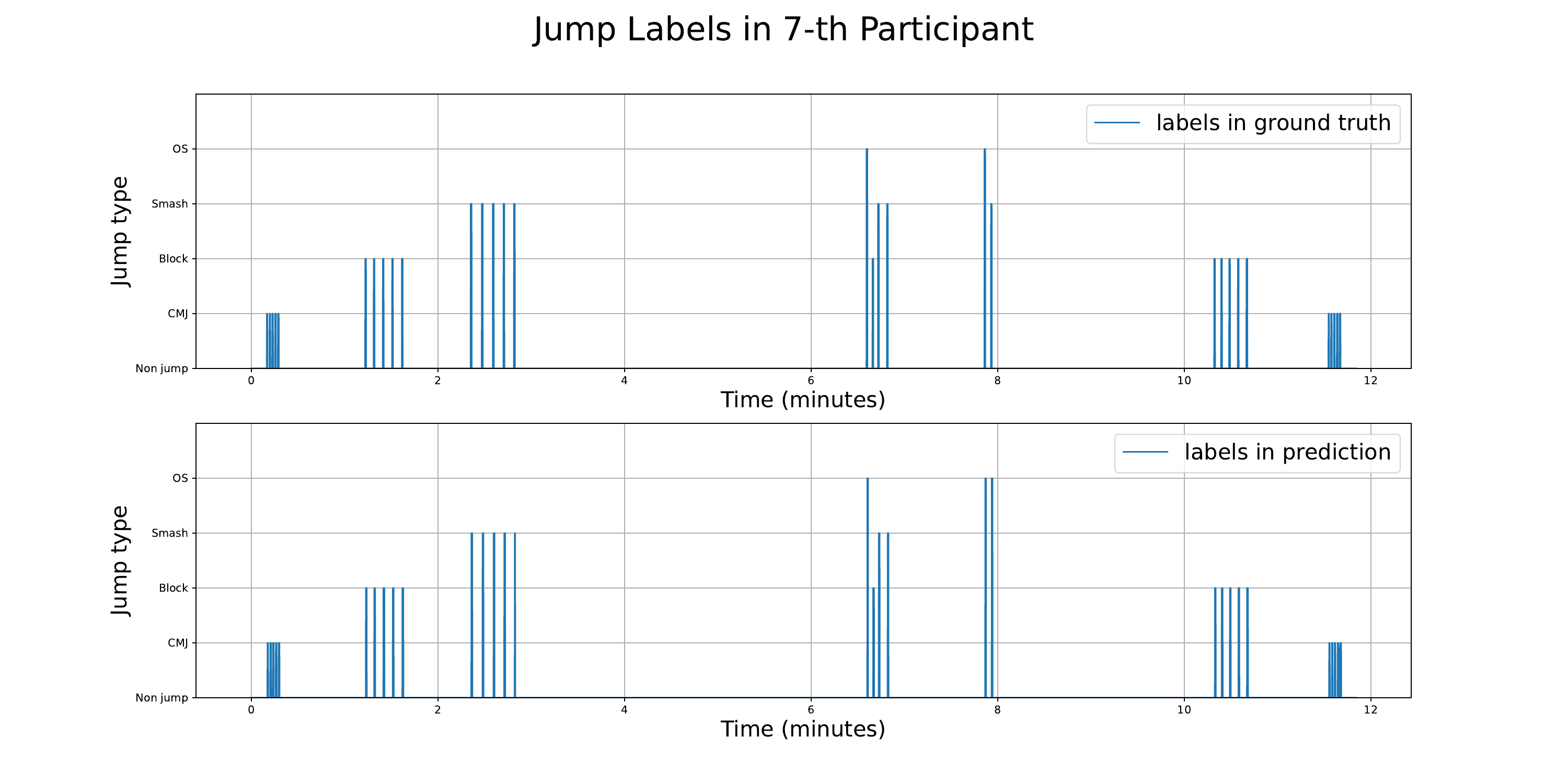}
    \caption{Difference of labels between video annotation and MS-TCN model output for 7th participant (the column represents different jump types)}
    \label{fig:3-2}
\end{figure}

\subsection{Height estimation}
Given that this study aims to predict peak jump height and that the dataset contains a limited number of jumps, we opted for ML models with simpler architectures than DL models. For model selection, we choose three classical ML models: RF, MLP and Extreme Gradient Boosting (XGboost), as demonstrated success in previous HAR research \cite{kanwal2023smartphone, rustam2020sensor}. Using the prepared ROI segments, we extract several time-domain and frequency-domain features using TSFEL package \cite{barandas2020tsfel}. A subset of 24 features, including maximum value, standard deviation, and spectral entropy from each signal channel, are selected based on prior research findings \cite{shang2024otago}. Additionally, the jump type is incorporated as one statistical feature. In total, 145 features are selected as the input to the regression model.

% In summary, the features used for height estimation are listed in Table \ref{tab: 2-1}.

% \begin{table}[htb]
%     \centering
%     \caption{Features used in height estimation task}
%     \begin{tabular}{c|c|c}
%     \hline
%        3-axis Accelerometer  & 3-axis Gyroscope & Meta features  \\
%     \hline
%        min\_acc  & min\_gyro  & jump\_classification\_label \\
%        max\_acc  & max\_gyro  & participant\_height \\
%        std\_acc  & std\_gyro  & participant\_weight \\

%        /  & /  & participant\_gender \\
%        /  & /  & participant\_dominant\_leg \\

%     \hline
%     \end{tabular}
%     \label{tab: 2-1}
% \end{table}

%% file: section/Experiments_and_discussion.tex
\subsection{Dataset preparation}
Following the experiment protocol as shown in the left part of Fig. \ref{fig:2-1}, participants were instructed to perform several common jumps or activities demanded in volleyball. Data were collected as a continuous time series for each participant. The volleyball-related activities include CMJ, Smash, Block, Overhead Serve (OS), Squat, and Dive. Note Squat and Dive were included considering their potential musculoskeletal loads although they are not jump activities. Besides, some other hops (jumps with extreme low heights) were included that release limited load on the patellar tendon. This study will not take Dive, Squat and other hops into physical load assessment, as the objective is to use jump height as an indicator for jumping tasks with high intensity. In total, the dataset includes 337 jumps and the duration of all recordings is 107.13 minutes.

Each participant was equipped with a waist-mounted IMU device (Actigraph Link GT9X) at a sampling frequency of 100 Hz. The channel number $C$ is set to 6 in this study, comprising three-axis accelerometer and three-axis gyroscope signals, as proven to be valuable in previous Human Activity Recognition (HAR) research \cite{shang2024ds, wang2024eating}. Additionally, one commercial device VERT was also worn for comparison. To obtain ground truth labels for each IMU signal sample, one camera (Sony Handycam HDR-PJ410) was used to record the video at 25 fps. Boris software (Behavioral Observation Research Interactive Software) was used for video annotation. Besides, video data from a 10-camera markerless motion capture system (Miqus, Qualisys, Sweden) was processed with Theia 3D software (Theia Markerless Inc., Kingston, ON, Canada), to obtain jump heights, which served as the gold standard in this study.

\begin{table}[t]
    \centering
    \caption{ML Models Parameters in Jump Height Estimation}
    \begin{tabular}{ c c c }
    \hline
     RF & MLP  & XGboost  \\
    \hline
    n\_estimators=50 & max\_iter=8000 & eta=0.1 \\
    max\_depth=10 & activation='relu' & n\_estimators=100   \\
    max\_leaf\_nodes=15 & solver='adam' & max\_depth=6 \\
    criterion='squared error' & / & gamma=0 \\
    \hline
    \end{tabular}
    \label{tab:3-1}
\end{table}

The data collection was approved by the ethical committee of the University Hospital in Ghent, Belgium (approval number BC-07679) \cite{shang2023multi}. All participants provided written informed consent prior to the experiments. Ten volleyball players were recruited for the study, including both elite and non-elite athletes. The only exclusion criterion was no recent history of lower limb pathology prior to the experiments. 

\subsection{Model training and evaluation methods}

We use the Leave-One-Subject-Out (LOSO) validation method to evaluate the model performance. During each iteration, nine out of ten participant data are used for training and the remaining one participant data is used for validation. The classification and regression model use the same group of data for training and testing respectively in each iteration of the pipeline.

For the jump classification model in the first stage, we input the raw IMU signals directly to the MS-TCN. The loss function for training MS-TCN includes Cross Entropy loss (CE) and Truncated Mean Squared Error (TMSE). CE is used for minimizing the classification difference between the ground truth and the prediction, whereas TMSE makes the predicted labels between adjacent samples consistent. The hyperparameter values of the loss function used in this study are the same as reported in \cite{farha2019ms}.

For the selection of baseline model, we choose the hybrid architecture including Convolutional Neural Networks (CNNs) and Long Short Term Memory (LSTM) for comparison, as CNN-LSTM is considered as the State-Of-The-Art (SOTA) in time-series prediction \cite{choudhury2023adaptive}. The parameters of both MS-TCN and CNN-LSTM are the same used in previous work \cite{farha2019ms, shang2023multi}. The MS-TCN and CNN-LSTM are implemented using Python PyTorch framework and executed on GPU P100.

\begin{table*}[h!]
    \centering
    \caption{LoA results of jump counts in prediction and annotation}
    \begin{tabular}{l l c c c c c }
    \hline
     &  & CMJ & Block & Smash & OS & All jumps \\
    \hline
    Ground truth & Jump counts & 109 & 113 & 82 & 33 & 337  \\
    \hline
    \multirow{2}{4em}{VERT} & Jump counts & / & / & / & / & 332 \\
    & LoA & / & / & / & / &  0.5 $\pm$ 1.39 \\
    \hline
    \multirow{2}{4em}{CNN-LSTM} & Jump counts & 126 & 118 & 100 & 40  & 384     \\
    & LoA & -1.7 $\pm$ 2.62  & -0.5 $\pm$ 8.37 & -1.8 $\pm$ 10.49 & -0.7 $\pm$ 8.62 & -4.7 $\pm$ 11.01 \\
    \hline
    \multirow{2}{4em}{MS-TCN} & Jump counts & 107 & 108 & 82 & 33 & 337     \\
    & LoA & \textbf{0.2 $\pm$ 0.83}  & \textbf{0.5 $\pm$ 4.16} & \textbf{0.1 $\pm$ 3.13} & \textbf{-0.8 $\pm$ 6.45}  & \textbf{0.0 $\pm$ 3.81} \\
    \hline
    \end{tabular}
    \label{tab:3-2}
    
\end{table*}

\begin{table*}[h!]
    \centering
    \caption{Segment-wise classification result on jump segments by CNN-LSTM and MS-TCN (Threshold = 0.1)}

    \begin{tabular}{l l c c c c c }
    \hline
     Method & Evaluation Metrics  & CMJ & Block & Smash & OS & All jumps  \\
    \hline
    \multirow{3}{4em}{CNN-LSTM} & Precision &  0.72 & 0.60 & 0.48 & 0.23 & 0.68   \\
     & Recall    &  0.80 & 0.60 & 0.57 & 0.28 & 0.75   \\
     & F1 score   & 0.75 & 0.60 & 0.52 & 0.25 & 0.71    \\
        
    \hline
     \multirow{3}{4em}{MS-TCN} & Precision & \textbf{0.88} & \textbf{0.85} & \textbf{0.87} & \textbf{0.45} & \textbf{0.89}    \\
     & Recall & \textbf{0.88} & \textbf{0.84} & \textbf{0.85} & \textbf{0.55} & \textbf{0.90}      \\
     & F1 score  & \textbf{0.88} & \textbf{0.85} & \textbf{0.86} & \textbf{0.49} & \textbf{0.90}     \\

    \hline
    \end{tabular}
    \label{tab:3-3}
\end{table*}

Theia 3D software estimates jump height based on the displacement of the Center of Mass (CoM) of the human body. In this dataset, Theia 3D successfully detects 320 out of 337 jumps, which serve as the gold standard. The features extracted from the corresponding IMU signal segments are then used to train the regression model. The model parameters for RF, XGBoost and MLP are listed in Table \ref{tab:3-1}. All models are implemented using Python scikit-learn library and executed on AMD Ryzen 7 PRO 8840U.

We use Limits of Agreements (LoA) to evaluate the classification model performance for jump occurrence. The definition of LoA is defined as $\mu_{diff} 
\pm 1.96 \times std_{diff}$, 
where $\mu_{diff}$ is the mean difference of the height between prediction and true values, and $std_{diff}$ is the standard deviation value of the difference. We compare the classification performance between MS-TCN with CNN-LSTM, also including the comparison with the commercial device VERT in jump count. To further evaluate the classification performance, we use the Intersection over Union (IoU), a commonly used metric for segment-wise evaluation \cite{wang2024evaluation}. By determining a threshold value, we can derive the number of True Positive (TP), False Positive (FP) and False Negative (FN), to compute Precision, Recall and F1 score for evaluation.

For the height estimation task, we compare the performance of three regression models against the VERT device to demonstrate that our method can outperform the existing commercial tools. The evaluation metrics include R squared score ($R^2$), Root Mean Square Error (RMSE), Mean Absolute Percentage Error (MAPE), and Pearson Correlation Coefficient (r). Additionally, we select the SHapley Additive exPlanations value (SHAP) for feature importance analysis to identify the most influential features extracted from six signal channels that contribute to height prediction \cite{lundberg2017unified}.

\subsection{Results and discussion}

\subsubsection{Jump type classification report}
As MS-TCN is trained for a sample-wise classification task, we have a predicted label for each timestamp within the time series data. We can observe the difference between video annotation and MS-TCN output as illustrated in Fig. \ref{fig:3-2} for one of the participants. Although few jumps are misclassified, most jumps can be accurately detected and classified.

\begin{figure}[htb]
    \centering
    \includegraphics[width=\linewidth]{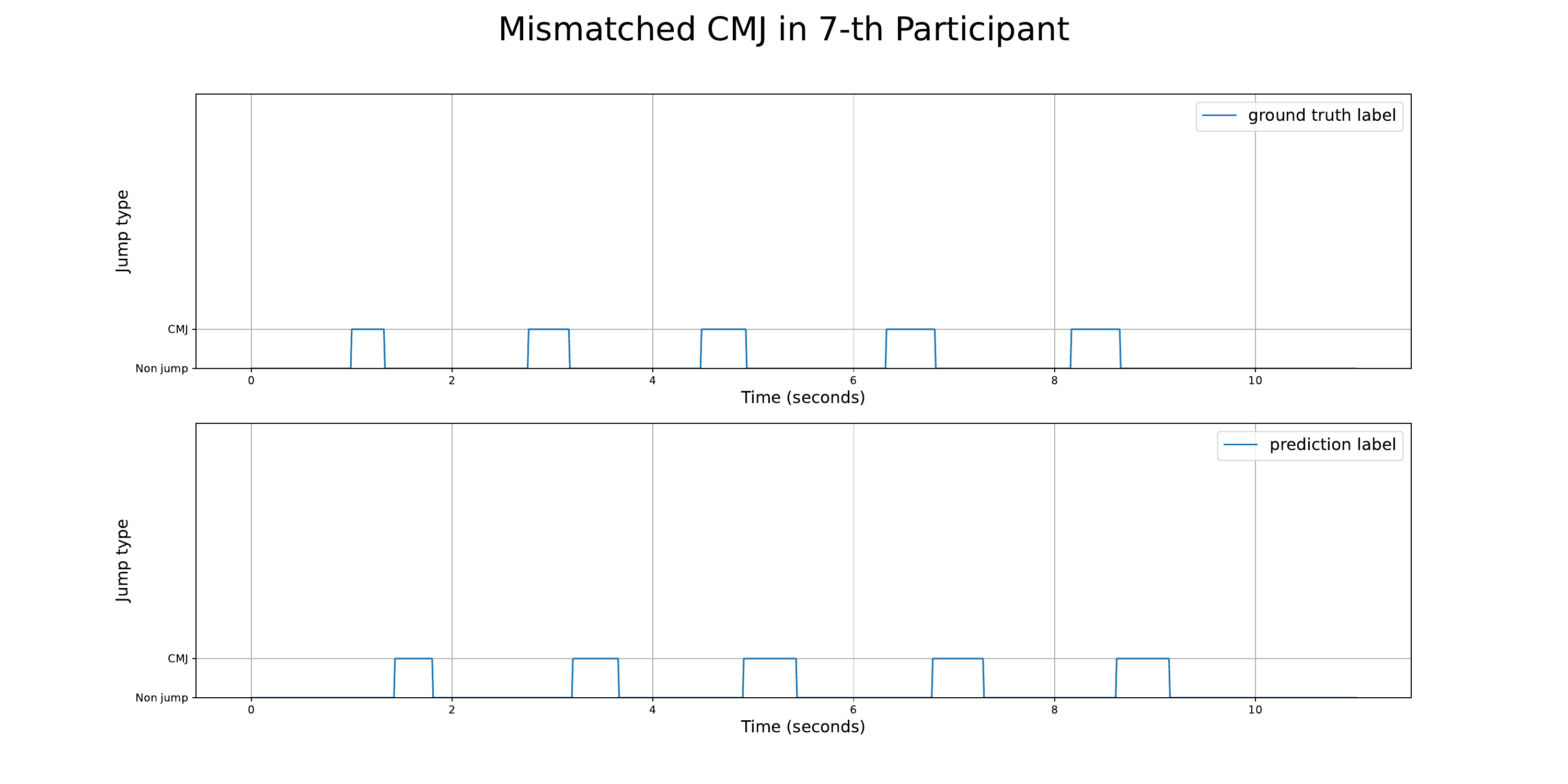}
    \caption{One example of mismatched labels (CMJ) between ground truth and prediction: predicted segment has a latency less than 0.5 second due to the error made in signal synchronization}
    \label{fig:3-3}
\end{figure}

The results of LoA are displayed in Table \ref{tab:3-2}. The LoA results show slight difference between each jump types and all jumps. Compared with CNN-LSTM, MS-TCN demonstrates a better performance, achieving an LoA of $0.0 \pm3.81$. The LoA of VERT is $0.5 \pm 1.39$, although it has a lower standard deviation, VERT tends to underestimate the actual number of jumps in this dataset. Moreover, VERT lacks the ability to classify jump types, which is crucial for performance enhancement and potential injury risk analysis \cite{migliorini2019injuries}. 

As the ultimate goal is to estimate jump heights based on the identified segments, we select a low threshold as $0.1$ in IoU to capture the most jumps, also accounting for potential mismatches caused by signal synchronization and differing sampling frequencies between devices. The IoU results are listed in Table \ref{tab:3-3}. We observe that both MS-TCN and CNN-LSTM struggle with accurately classifying the OS jump type. This can be attributed to the similarities in jump patterns between OS, Smash, and Block in the controlled lab environment, where all involve players stepping forward and jumping above the test area. However, in real-game settings, these jumps serve distinct purposes. Smash is the most common jump type used for attacking and Block is primarily defensive near the net, whilst OS is typically performed at the start of the round as the first attack to pass the volleyball over the net.

By analyzing the signals associated with missing jumps, we attribute the issue to annotation errors or inaccurate recording timestamps. As shown in Fig. \ref{fig:3-3}, we observe there is a 0.5-second latency between the predicted and annotated jump events. Although our method successfully detects these mismatched jumps according to the jump count results listed in Table \ref{tab:3-2}, we only select the segments marked as TP by IoU to evaluate the regression model. This study serves as an initial validation of the feasibility of an AI-assisted pipeline for jump height estimation. To extend our method to a broader range of jump activities in future work, precise data alignment will be crucial.

\begin{figure}[t]
    \centering
    \begin{subfigure}[t]{\columnwidth}
        \includegraphics[width = \linewidth]{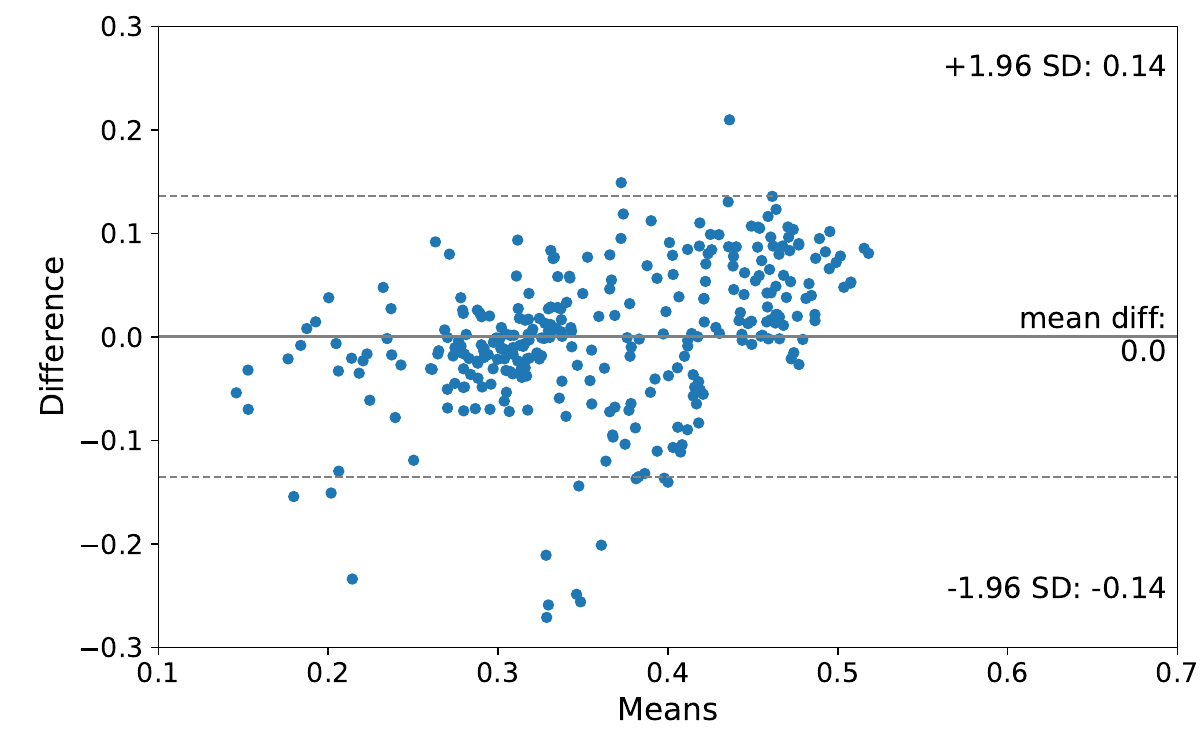} \caption{Bland–Altman plot of RF and Theia 3D}
    \end{subfigure}
    \begin{subfigure}[t]{\columnwidth}
        \includegraphics[width = \linewidth]{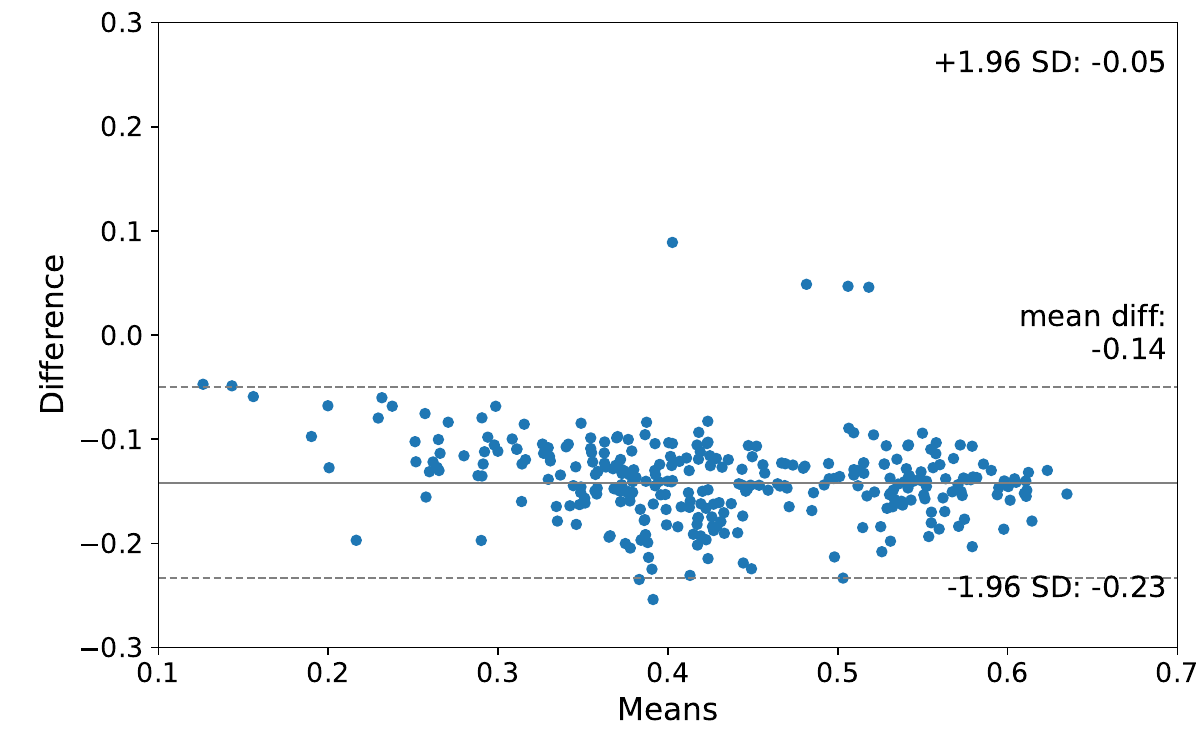}
        \caption{Bland–Altman plot of VERT and Theia 3D}
    \end{subfigure}

    \caption{Bland–Altman plot of jump height (m) estimation by RF (above) and VERT (below)}
    \label{fig:3-4}
\end{figure}

\subsubsection{Jump height estimation report}
The regression model is trained using jump segments identified from video annotation and Theia 3D, capturing 320 out of 337 total jumps. Testing is conducted on segments classified as TP in the classification report. As shown in Table \ref{tab:3-3}, approximately 90\% jumps are successfully detected by MS-TCN, with 279 jumps used for height estimation. Four additional jumps are excluded as they are not captured by the VERT device. To ensure a fair comparison, performance evaluation is limited to jumps detected by both methods.

 According to the results listed in the above part of Table \ref{tab:3-4}, the three models yield comparable results, with the RF model achieving the best performance with a $R^2$ value of 0.50. Although VERT performs well for jump counting, it fails to provide reliable height estimation in the complex real-game settings we studied ($R^2 = -1.53$). Due to the annotation error mentioned in Fig. \ref{fig:3-3}, not all predicted jump segments could be classified as TP for use in the pipeline, which can be solved with more precise data alignment. In the below part of Table \ref{tab:3-4}, all jumps captured by VERT are used for evaluation (315 jumps in total), with video-annotated segments directly input into the height prediction model. The observed higher value of $R^2$ and r, indicate reduced discrepancies between predictions and the gold standard. These findings reinforce the feasibility of AI-assisted approaches for estimating heights.

\begin{table}[t]
    \centering
    \caption{Comparison results of our method and VERT device: the above part is calculated on the (275) jumps identified by both MS-TCN and VERT device, and the below part is calculated on the (315) jumps identified by VERT and video annotation.}
    \begin{tabular}{l l c c c c}
    \hline
     & Methods &  MAPE  & RMSE (m) & $R^2$ & r \\
    \hline
     & VERT & 0.43 & 0.15 & -1.53 & 0.89 \\
    \hline
    \multirow{3}{6.5em}{Segments from MS-TCN} & RF  & 0.15 & 0.07 & 0.50 & 0.71 \\
    & XGBoost  & 0.16 & 0.07 & 0.45 & 0.68 \\
    & MLP  & 0.16 & 0.07 & 0.48 & 0.72 \\
    \hline
    \hline
    & VERT  & 0.44 & 0.15 & -1.19 & 0.90 \\
    \hline
    \multirow{3}{6.5em}{Segments from Video annotation} & RF  & 0.16 & 0.07 & 0.53 & 0.73 \\
    & XGBoost   & 0.18 & 0.07 & 0.48 & 0.70 \\
    & MLP  & 0.17 & 0.07 & 0.52 & 0.75 \\
    \hline
    \end{tabular}
    \label{tab:3-4}
\end{table}

\begin{figure}[t]
    \centering
    \includegraphics[width=\linewidth]{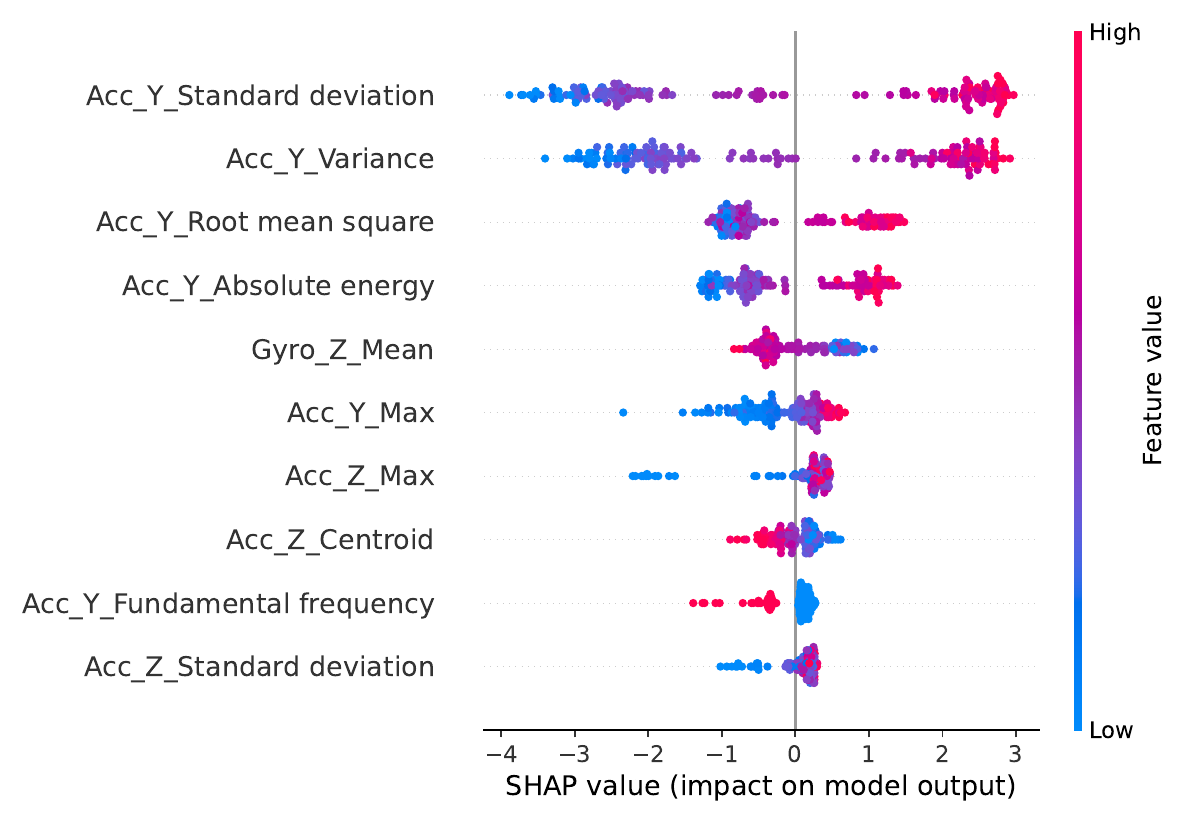}
    \caption{Top ten important features calculated by SHAP method: the color indicates the magnitude of feature value, and the SHAP value denotes the importance of the feature to the prediction (jump heights)}
    \label{fig:3-5}
\end{figure}

The Bland-Altman plot in Fig. \ref{fig:3-4} compared the difference between predicted and actual jump heights by our method (RF) and VERT. The plot of our proposed method reveals that most outliers occur at specific jump heights, suggesting that limited data can lead to poorer performance. The LoA area indicates the mean difference between predicted height and ground truth is close to zero, with an upper/lower bound of 0.14 (14 centimeters) within an acceptable range. Although VERT method has a smaller standard deviation value, it tends to overestimate the jump heights in most cases, resulting in poor performance as shown in Table \ref{tab:3-4}.

The SHAP values for the top 10 important features are illustrated in Fig. \ref{fig:3-5}. The result is computed based on the RF model. The plot indicates the signals along the  vertical axis (y-axis) and sagittal axis (z-axis) have a larger impact than frontal axis (x-axis). Additionally, variations in jump patterns, such as with lower standard deviation in y-axis acceleration, can increase the difficulty of prediction. Future work can consider training the model more diverse datasets to improve robustness and accuracy.